\documentclass[useams,usenatbib]{mn2e}
\usepackage{graphicx}

\title[Near-IR spectroscopy of a young SSC in NGC~6946]{Near-IR  
spectroscopy of a young super-star cluster in NGC~6946:    
chemical abundances and abundance patterns
\thanks{Based on data obtained at the W.M.Keck Observatory,
        which is operated as a scientific partnership among the California
        Institute of Technology, the University of California, and the National
        Aeronautics and Space Administration. The Observatory was made possible
        by the generous financial support of the W.M. Keck Foundation.}}

\author[Larsen et al.]{S. S. Larsen$^1$, L. Origlia$^2$, J. P. Brodie$^3$,
  J. S. Gallagher, III$^4$.\\
 $^1$ ESO / ST-ECF, Karl-Schwarzschild-Str.\ 2, D-85748 Garching bei
      M{\"u}nchen, Germany, e-mail slarsen@eso.org \\
 $^2$ INAF-Osservatorio Astronomico di Bologna, Via Ranzani 1, I-40127 Bologna,
      Italy, e-mail livia.origlia@bo.astro.it \\
 $^3$ UCO/Lick Observatory, 1156 High Street, University of California, 
      Santa Cruz, CA 95064, USA, email brodie@ucolick.org \\
 $^4$ Astronomy Department, University of Wisconsin, 475 North Charter
      Street, Madison, WI 53706, USA, email jsg@astro.wisc.edu
       }
\date{Accepted 2006 January 16. Received 2006 January 16; in original form 2005 December 22.}

\begin{document}
\pagerange{\pageref{firstpage}--\pageref{lastpage}} \pubyear{2006}
\maketitle
\label{firstpage}

\begin{abstract}
Using the NIRSPEC spectrograph at Keck II, we have obtained $H$ and $K$-band
echelle spectra for a young ($\sim10-15$ Myr), 
luminous ($M_V\sim-13.2$) super-star cluster in the nearby spiral galaxy
NGC~6946.  From spectral synthesis and equivalent width measurements 
we obtain for the first time accurate abundances and abundance patterns 
in an extragalactic super-star cluster.  We find [Fe/H]$=-0.45\pm0.08$~dex,
an average $\alpha$-enhancement of $\approx+0.22\pm0.1$~dex, and a
relatively low  $\rm ^{12}C/^{13}C\approx 8\pm2$ isotopic ratio. 
We also measure a velocity 
dispersion of $\approx$9.1~km/s, in agreement with previous estimates. 
We conclude that integrated high-dispersion spectroscopy of massive
star clusters is a promising alternative to other methods for abundance 
analysis in extragalactic young stellar populations.
\end{abstract}

 
\section{Introduction}

Star clusters have a long history as important tools for addressing a wide
range of questions in astronomy. With few exceptions, they are ``simple
stellar populations'' (SSPs), i.e.\ they are composed of stars born in a 
single burst and sharing the same chemical composition (at least to 
first order) and age. They have played an important role as test labs for 
models of stellar evolution \citep[e.g.][]{mm81,ren88,chi92}. As confidence has
grown in our ability to model their integrated properties, so has 
their importance as tracers of stellar populations in galaxies that are 
too distant for individual stars to be resolved.

The latter point is perhaps best illustrated by considering, for the moment, 
the old \emph{globular clusters} (GCs) which are ubiquitous in all major 
galaxies.  GCs typically contain large numbers of stars ($10^4 - 10^6$),
so most phases of stellar evolution are well sampled and stochastic effects
therefore minimized. Although there are still unsolved problems (e.g.\
concerning horizontal branch morphology), SSP models can
now provide a reasonably realistic account of integrated GC
observables such as broad-band colours and absorption line strengths as a 
function of
age and metallicity.  Thanks to the availability of efficient spectrographs
on 8-10 m class telescopes, spectroscopy of GCs in galaxies well beyond the
Local Group is now routinely feasible, and has been utilized in many
studies to probe the stellar populations in early-type galaxies and 
constrain their ages and metallicities 
\citep[e.g.][and references therein]{bs06}.
Much of this work has relied on measurements of absorption line features
at relatively low ($\sim10$ \AA) spectral resolution. In the optical, the 
Lick/IDS system of absorption line indices has found wide-spread use 
\citep{bur84,tra98}, while indices for the near-infrared have been defined by
\citet{ba87} and \citet{vaz03}.  Ages are typically estimated from Balmer 
line strengths (mainly H$\delta$, H$\gamma$, and H$\beta$), with other
indices being more sensitive to metallicity. However, detailed abundances
of individual elements typically cannot be reliably measured at this 
resolution. In addition, the Lick/IDS system is not well tailored for studying
young stellar populations, partly due to limitations in the empirical 
libraries used in the construction of SSP models, but also because the index
definitions are designed primarily for studies of old stellar populations.

Until now, the main source of information about the chemical composition of 
\emph{young} stellar populations beyond the Local Group has been
measurements of emission lines in H{\sc ii} regions. However, these provide 
access to only a limited set of elements (mainly O, N, S) and often have to 
rely on empirical calibrations of line strengths vs.\ metallicity. The auroral 
lines (e.g.\ [O{\sc iii}] 4363\AA) are usually too faint to be measured 
directly, 
thus prohibiting a determination of the nebular temperature \citep{sta01}.  
Furthermore, H{\sc ii} regions only offer a snapshot of the \emph{present-day} 
chemistry and do not provide any information about the past history.

  Spectroscopy of (massive) star clusters has the potential to provide 
information on 
the entire star formation histories of galaxies.  For masses in the $10^5$ 
M$_\odot$ -- $10^6$ M$_\odot$ range, star clusters typically have velocity 
dispersions of about 5--10 km/s, allowing studies of their integrated 
properties at spectral resolutions up to $\lambda/\Delta\lambda \approx$
30,000 or more.  This is sufficient to constrain individual element abundances.
However, extending this type of analysis to young clusters comes with its own 
set of difficulties.  Compared to studies of old GCs, some of the main 
challenges
are: 1) models for the massive stars found in clusters with ages of a few
$\times10^7$ years are much less certain than those of the low-mass stars
found in old GCs \citep{mas03} 2) the
optical spectra are generally dominated by the relatively featureless
spectra of hot stars, diluting absorption features and making them
harder to measure, 3) for all but the most massive clusters, the integrated
spectra may be dominated by a few massive stars, causing unpredictable
stochastic fluctuations in integrated properties \citep[e.g.][]{lm00}.

In this letter we aim to take a first step towards deriving abundances
for \emph{young} extragalactic star clusters from their integrated light,
by modelling the near-infrared spectrum of a luminous ($M_V=-13.2$) 
young ($\sim10$ Myr) star cluster in the nearby
spiral galaxy NGC~6946. This object was first identified as a star
cluster by \citet{lr99} and was labelled NGC6946-1447 by them, but is located 
within a larger stellar complex first noted by \citet{hod67}.  Both
$UBVI$ broad-band colours and spectroscopic observations of Balmer and
He I absorption lines yield consistent age estimates in the range
10--15 Myr \citep{laretal01,efre02}.
From the luminosity and estimated age of the cluster, SSP
models yielded a mass of $\sim0.8\times10^6$ M$_\odot$
for a Salpeter-like IMF extending down to 0.1 M$_\odot$, or 
$\sim0.55\times10^6$ M$_\odot$ if the IMF is log-normal
below 0.4 M$_\odot$ \citep{laretal01}.  A dynamical mass estimate, based 
on high-dispersion optical spectroscopy and HST imaging, yielded a
somewhat higher mass of $\sim1.7\times10^6$ M$_\odot$. Such massive clusters
are sometimes called ``super star clusters'' (SSCs).
Using isochrones from \citet{gir00}, we estimate that the cluster contains
about 130 red supergiants.
A detailed description of the cluster and 
surrounding stellar complex is given in \citet{laretal01,lar02}.

No line emission is observed from the cluster itself but \citet{efre02}
estimated an oxygen abundance of $12 + \log(O/H) = 8.95 \pm 0.2$ from
long-slit spectroscopy of nearby H{\sc ii} regions.
\citet{br92} measured O abundances for
H{\sc ii} regions distributed throughout the disk of NGC~6946 (using
narrow-band imaging of O, N and H emission lines) and derived
an oxygen abundance gradient of 
$\Delta \log (O/H) / \Delta R = -0.089 \pm 0.003 \, {\rm dex}\,  {\rm kpc}^{-1}$ 
and a central value of $12 + \log (O/H)$ = 9.37. At the projected
galactocentric distance of NGC6946-1447 (4.8 kpc), this corresponds to
$12 + \log (O/H) = 8.94 \pm 0.01$.  \citet{kob98} quote an O abundance
of $12 + \log (O/H) = 9.13$ at a radius of 3 kpc, or $12 + \log (O/H) = 8.97$ 
at 4.8 kpc if we adopt the abundance gradient from \citet{br92}.
Thus, all available measurements consistently give $12+\log(O/H)$ 
between 8.94 and 8.97,
although the above O abundances are all based on measurements of the 
collisionally excited O lines vs.\ Balmer line ratios and may therefore be 
subject to systematic uncertainties at the 0.1-0.2 dex level.

\section{Observations and Data Reduction}


\begin{figure*}
\centering
\includegraphics[width=14.2cm]{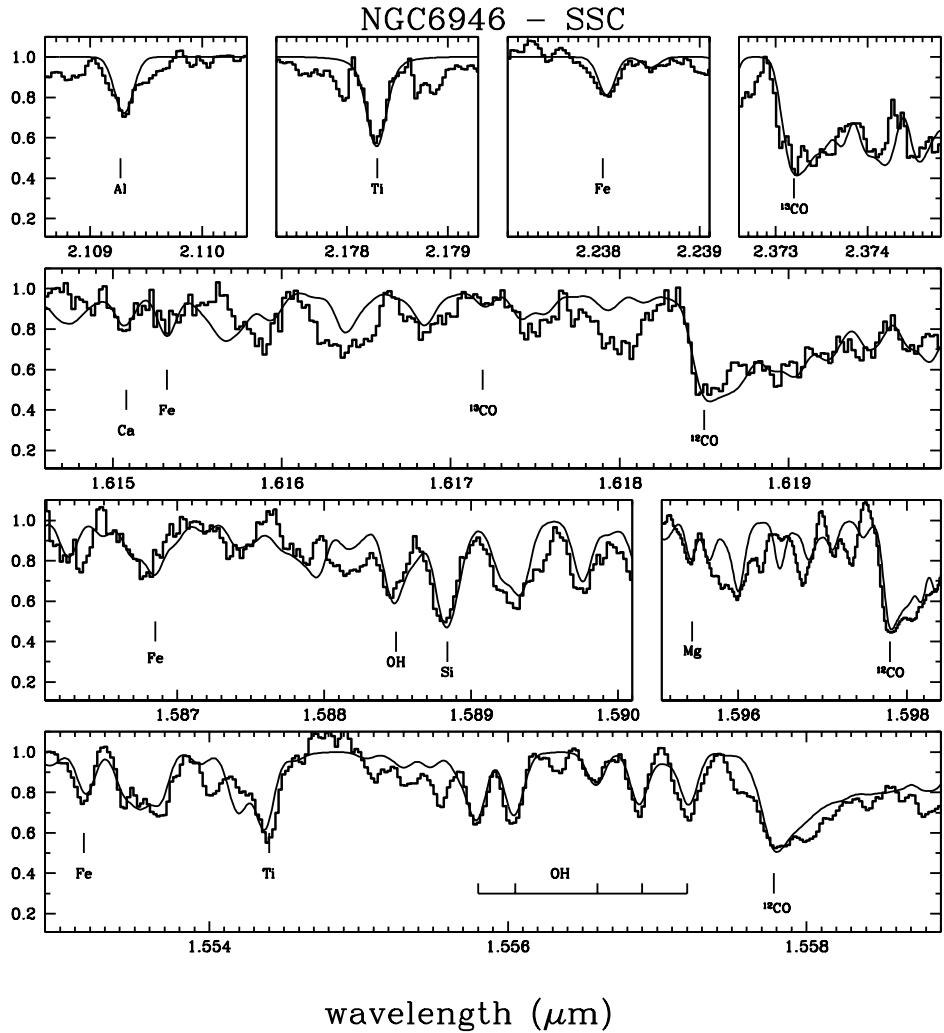}
\caption {Near-IR spectra of the SSC of NGC~6946.
Observed spectra: histograms; synthetic stellar best-fit solution: 
solid lines. A few atomic and molecular features of interest are also marked.
\label{IR}}
\end{figure*}

H and K-band high-resolution spectra 
were acquired on 13 July 2002, using the infrared spectrograph NIRSPEC 
\citep{ml98} mounted at the Nasmyth focus of the Keck~II telescope.
The high resolution echelle mode, with a slit width of $0\farcs43$
(3 pixels) and a length of 12\arcsec\ was used, providing a
spectral resolution of $\lambda/\Delta\lambda = 25,000$.  The exposure times 
were 64 min and 48 min in the H and K bands, respectively, yielding
a S/N of about 26 and 36 per pixel in the dispersion direction
(the $K$ magnitude of the cluster is about 13.0).
The observations were obtained in pairs of exposures with a duration of
240 s each, nodded a few arcsec along the slit to allow
reliable sky subtraction without any additional overhead for
separate sky exposures. 
The NIRSPEC data has previously been used by \citet{lbh04} to 
derive a line-of-sight velocity dispersion of $8.8\pm1.5$ km/s for the 
cluster, in good agreement with the value of $10.0\pm2.7$ km/s
derived by \citet{laretal01} (based on Keck/HIRES spectroscopy). We
refer to \citet{lbh04} for details on the data reduction.


\section{Spectral Analysis}
\label{analysis}

Near-IR spectroscopy is a powerful tool to obtain accurate abundances of 
key metals in cool stars ($\rm T_{\rm eff}\le 5000$~K).  Several atomic and 
molecular lines are strong and not affected by severe blending, 
making them powerful abundance tracers not only in stars but also in more
distant stellar clusters and galaxies and for a wide range of metallicities 
and ages \citep{ori97,oo98,ori04}.  However, to properly account for line 
blending, abundance analysis from integrated spectra generally still requires 
full spectral 
synthesis techniques and not just equivalent width measurements of individual 
lines.  Population synthesis may also be required to define the dominant 
contribution to the stellar luminosity.

The near IR stellar continuum of young stellar clusters and starburst galaxies  
is almost entirely due to luminous red supergiants 
\citep{oo00} and usually it dominates over nebular and dust emission.
Based on the Girardi et al.\ isochrones, stars hotter than
10,000 K contribute only $\sim5$\% of the $H$- and $K$-band flux at 15 Myrs.
This represents a major, conceptual simplification in population and
spectral synthesis techniques, making the interpretation
of the integrated spectra much easier.  The spectra can be modelled with
an equivalent, average star, whose stellar parameters
(temperature T$_{\rm eff}$, gravity log{\it g} and microturbulence 
velocity $\xi$) mainly depend on the stellar age and metallicity.
Both observations and evolutionary models
\citep[see e.g.][and references therein]{k99,o99,mas03}
suggest that red supergiants of ages between $\simeq $6 and 100 Myr and
metallicities between 1/10 and Solar have low gravities
(log~g$<$1.0), low temperatures ($\le$ 4000~K)
and relatively high microturbulence velocity ($\xi\ge$3 km/s). 


\begin{table*}
\begin{center}
\caption{Adopted stellar atmosphere parameters and abundance estimates for the SSC in NGC~6946.
\label{ab}}
\begin{tabular}{lccccccccccc}
\hline\hline
T$_{\rm eff}$ [K] &
log~g &
$\xi$ [km~s$^{-1}$]   &
$\rm [Fe/H]$   &
$\rm [O/Fe]$   &
$\rm [Ca/Fe]$   &
$\rm [Si/Fe]$   &
$\rm [Mg/Fe]$   &
$\rm [Ti/Fe]$   &
$\rm [\alpha/Fe]^a$&
$\rm [Al/Fe]$   &
$\rm [C/Fe]$   \\
\hline
4000 & 0.5 & 3 & -0.45 & 0.28 & 0.25 & 0.07 & 0.25 & 0.30 & 0.22 & 0.25 & -0.25\\  
& & &$\pm$0.08&$\pm$0.09&$\pm$0.12&$\pm$0.18&$\pm$0.17&$\pm$0.11&$\pm$0.11&$\pm$0.18&$\pm$0.11\\
\hline
\end{tabular}
\end{center}
$^a$ $\rm [\alpha/Fe]$ is the average $\rm [<Ca,Si,Mg,Ti>/Fe]$ abundance ratio.
\end{table*}


At the NIRSPEC resolution of R=25,000, several single roto-vibrational OH 
lines and CO bandheads can be measured and used to derive accurate oxygen and 
carbon abundances.  Although our NIRSPEC setup was optimised for velocity
dispersion measurements rather than abundance 
analysis, abundances of other metals can be derived from the atomic 
lines of Fe~I, Mg~I, Si~I, Ti~I, Ca~I and Al~I.  

A grid of synthetic spectra of red supergiant stars for different input 
atmospheric parameters and abundances were computed, using an updated 
\citep{ori02,ori03} version of the code described in \citet{ori93}.
Briefly, the code uses the LTE approximation and is based on molecular 
blanketed model atmospheres of \citet{jbk80} at temperatures $\le$4000~K
and the ATLAS9 models for temperatures above 4000~K.
Recently, the NextGen model atmospheres \citep{hau99} have been also
implemented within the code and tested. Compared with the 
older models, the differences in the resulting abundances are only
minor \citep[well within a few hundredths dex;][]{ro05}.  This is not 
surprising, since the major source of opacity in the near IR spectra
of cool stars is H$^-$ with a minimum around 1.6$ \mu$m and 
small differences in the temperature structure
of different model atmospheres have a minor impact on the overall 
abundance determination.
The code also includes several thousands of near IR atomic lines and molecular
roto-vibrational transitions due to CO, OH and CN.
Three main compilations of
atomic oscillator strengths are used, namely
the Kurucz's database
(c.f. {\it http://cfa-www.harward.edu/amdata/ampdata/kurucz23/sekur.html}),
and those published by \citet{bg73} and \citet{mb99}.

The code provides full spectral synthesis over the 1--2.5 $\mu$m range
and abundance estimates are mainly obtained by best-fitting the full
observed spectrum and by measuring the equivalent widths
of a few selected features (Fig.~\ref{IR}), dominated by
a specific chemical element, as a further cross-check.
The equivalent widths were measured by performing a Gaussian fit
with $\sigma$ equal to the measured stellar velocity dispersion;
typical values range between 100 and 500 m\AA\ with a conservative error 
of $\pm$20~m\AA\ to also account for a $\pm$2\%
uncertainty in the continuum positioning.
By best-fitting the full observed IR spectrum and by measuring the equivalent
widths of selected lines,
we obtained the following stellar parameters and abundance
patterns:
T$_{\rm eff}$=4000, log{\it g}=0.5, $\xi$=3,
[Fe/H]=--0.45; [O/Fe]=+0.28, [$<$Si,Mg,Ca,Ti$>$/Fe]=+0.22;
[Al/Fe]=+0.25; [C/Fe]=--0.25.
We also measure $\rm ^{12}C/^{13}C\approx8\pm2$.

Table~\ref{ab} lists the derived abundances and their associated random errors
at 90\% confidence. Reference Solar abundances are from \citet{grev98}.
In addition, we also measure a heliocentric radial velocity of 
$\rm<v_r>=+141\pm 2~km/s$  
and velocity dispersion (corrected for instrumental broadening) of 
$\approx9.1\pm1$ km/s.  These numbers agree reassuringly well with previous
estimates \citep{efre02,laretal01,lbh04}.

\subsection{Error budget}

\begin{figure}
\centering
\includegraphics[width=7.5cm]{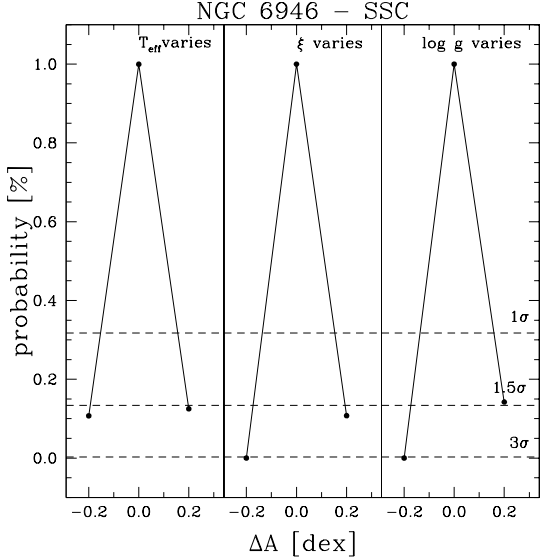}
\caption{
Average probability of a random realization of our best-fitting solution and 
the test models with varying temperature by $\Delta T_{\rm eff}$ of 
$\pm$200K (left panels), microturbulence by $\Delta \xi$ of $\mp$1.0 km 
s$^{-1}$ (right panels),  and gravity by $\Delta log~g$ of $\pm$0.5 dex 
(middle panels), with respect to the best-fitting solution (see 
Sect.~\ref{analysis}) for the SSC.
\label{IRtest}}
\end{figure}

Synthetic spectra with lower element abundances are \emph{systematically}
shallower (have weaker features) than the best-fit solution, while the 
opposite occurs when higher abundances are adopted.  In order to check the 
statistical significance of our best-fit solution, as a function of merit 
we adopt the difference between the model and the observed spectrum 
(hereafter $\delta$).  This parameter is more effective for quantifying
systematic discrepancies than the classical $\chi^2$ test, 
which is equally sensitive to \emph{random} and \emph{systematic} scatters 
\citep[see][for more details]{ori03}.

Since $\delta$ is expected to follow a Gaussian distribution, we compute 
$\overline{\delta}$ and the corresponding standard deviation ($\sigma$) for 
the best-fit solution and 6 {\it test models} with stellar parameters varying 
by $\Delta $T$_{\rm eff}$$\pm$200~K,
$\Delta $log~g=$\pm$0.5 and $\Delta \xi$$\mp$1.0~~km~s$^{-1}$ with respect 
to the best-fit, and abundances varying accordingly by $\approx\pm0.2$~dex,
in order to still reproduce the depth of the observed features.  We then 
extract 10000 random subsamples from each {\it test model} (assuming a 
Gaussian distribution) and we compute the probability $P$ that a random 
realization of the data-points around a {\it test model} display a 
$\overline{\delta}$ that is compatible with the {\em best-fit} model.
$P\simeq 1$ indicates that the model is a good representation of the
observed spectrum.

The left panel of Fig.~\ref{IRtest} shows the results for the
observed H band spectrum of the SSC in NGC~6946.
It can be easily appreciated that the best-fit solution provides
in all cases a clear maximum in $P$ ($>$99\%)
with respect to the {\it test models},
which are statistical significant only at 
$>1.5\sigma$ level. 
We also computed {test models} with the same stellar parameters 
as the best-fit solution and varying only the abundances.
Models with $\pm 0.1$~dex with respect to the best-fit 
solution are significant at $1-1.5 \sigma$ level, while models 
with $\pm 0.2$~dex are only marginally acceptable at a $>3 \sigma$ level.

Hence, as a conservative estimate of the systematic error in the derived 
best-fit abundances,
due to the residual uncertainty in the adopted stellar parameters, one can
assume a value of $\approx \pm 0.1$~dex. 
Moreover, since the
stellar features under consideration show a similar trend
with variation in the stellar parameters, although with different
sensitivity, {\it relative } abundances are less
dependent on stellar parameter assumptions, 
reducing the systematic uncertainty to $<$0.1~dex. 

\section{Discussion and Conclusions}

Based on near-infrared $H$ and $K$-band spectroscopy, we have carried out
a detailed abundance analysis of a young massive star cluster in the
nearby spiral galaxy NGC~6946. 
In this exploratory work, we have derived abundances of several individual
key elements, including Fe, Ca, Si, Mg and Ti. We find a sub-solar Fe abundance
([Fe/H] = $-0.45\pm0.08$), while the 
$\alpha$-element to Fe abundance ratios are all enhanced with
respect to the Solar values with a mean [$\alpha$/Fe] = $0.22\pm0.11$.
The O abundance derived from the cluster spectrum is [O/H]$ = -0.17\pm0.09$,
about 0.3 dex lower than the value based on H{\sc ii} regions at the same 
galactocentric distance.
We find a $^{12}$C/$^{13}$C ratio of $\approx8\pm2$, similar to or slightly
lower than typically observed in red supergiants in Galactic open 
clusters \citep{luck94,gw00} and in the (more metal-poor) 
cluster NGC~330 in the SMC \citep{gw99}.  Standard stellar models 
\citep{schal92} predict a surface $^{12}$C/$^{13}$C ratio of $\sim17$ for a 
15 M$_\odot$, $\sim12$ Myr-old (Solar metallicity) red supergiant, about a 
factor of two higher than the value derived here.

Super-solar [$\alpha$/Fe] ratios are usually interpreted as signatures
of rapid, bursty star formation, with the gas mainly enriched by
Type II supernovae with short-lived, massive progenitor stars 
\citep{mac97}. Stellar populations formed over timescales of several Gyrs or 
more are expected to show solar-like $\alpha$/Fe-element abundance
ratios, as observed in the Milky Way thin disk.

In this context, it is worth noting that the complex hosting NGC6946-1447 
may qualify as a `localized starburst' \citep{efre04}.  Timing is a 
critical issue, however: in order to produce 
super-solar [$\alpha$/Fe] ratios, the starburst must have \emph{preceded}
the formation of the cluster itself by several Myrs. The delay must have
been long enough for Type II SNe to produce significant amounts of
$\alpha$-elements, and enough time must then have elapsed to allow
mixing of the pre-existing gas with $\alpha$-enhanced ejecta before the 
cluster formed. Interestingly, reconstruction of the field star
colour-magnitude diagram has provided some evidence for star 
formation in the complex at least 10--15 Myr prior to the formation of the 
cluster \citep{lar02}.  Alternatively, the current global
SFR in NGC~6946 may be sufficiently elevated above the past average to
cause a general enrichment of the ISM with $\alpha$-elements. Addressing 
these questions more
quantitatively would require a detailed modeling of the chemical
enrichment history and a knowledge of the past star formation history
which is currently not available.

This study represents one of the first cases of a detailed abundance
analysis of a star-forming galaxy beyond the Local Group.  While many
uncertainties remain, we suggest that observations of extragalactic young 
star clusters hold great potential for constraining the chemical enrichment 
histories of their parent galaxies. 

\section*{Acknowledgments}

JPB acknowledges support from NSF grant AST-0206139.
We acknowledge the Keck Observatory and the NIRSPEC team.
The authors wish to recognize and acknowledge the very significant cultural
role and reverence that the summit of Mauna Kea has always had within
the indigenous Hawaiian community.
We are most fortunate to have the opportunity to conduct observations 
from this mountain.

\label{lastpage}


\begin{thebibliography}{}

\bibitem[\protect\citeauthoryear{Belley \& Roy}{1992}]{br92}
  Belley, J., \& Roy, J.-R., 1992, ApJS, 78, 61

\bibitem[\protect\citeauthoryear{Bica \& Alloin}{1987}]{ba87}
  Bica, E., \& Alloin, D., 1987, A\&A, 186, 49

\bibitem[\protect\citeauthoryear{Bi\`emont \& Grevesse}{1973}]{bg73}
Bi\`emont, E., \& Grevesse, N. 1973, {\it Atomic Data and Nuclear Data
Tables}, 12, 221

\bibitem[\protect\citeauthoryear{Brodie \& Strader}{2006}]{bs06}
Brodie, J.\ P., \& Strader, J., 2006, ARA\&A, in prep.

\bibitem[\protect\citeauthoryear{Burstein et al.}{1984}]{bur84}
Burstein, D., Faber, S. M., Gaskell, C. M., \& Krumm, N., 1984, ApJ, 287, 586

\bibitem[\protect\citeauthoryear{Gonzalez \& Wallerstein}{1999}]{gw99}
  Gonzalez, G., \& Wallerstein, G., AJ, 117, 2286

\bibitem[\protect\citeauthoryear{Gonzalez \& Wallerstein}{2000}]{gw00}
  Gonzalez, G., \& Wallerstein, G., AJ, 119, 1839

\bibitem[\protect\citeauthoryear{Chiosi et al.}{1992}]{chi92}
  Chiosi, C., Bertelli, G., \& Bressan, A., 1992, ARA\&A, 30, 235

\bibitem[\protect\citeauthoryear{Efremov}{2004}]{efre04}
  Efremov, Yu. N., 2004, contribution to the transactions of the
  conference 'Gamov-100', Odessa, August 2004, astro-ph/0410702

\bibitem[\protect\citeauthoryear{Efremov et al.}{2002}]{efre02}
  Efremov, Yu. N., Pustilnik, S. A., Kniazev, A. Y., Elmegreen, B. G.,
  Larsen, S. S., Alfaro, E. J., Hodge, P. W., Pramsky, A. G., \& Richtler, T.,
  2002, A\&A 389, 855

\bibitem[\protect\citeauthoryear{Girardi et al.}{2000}]{gir00}
 Girardi, L., Bressan, A., Bertelli, G., Chiosi, C.\ 2000, A\&AS, 141, 371

\bibitem[\protect\citeauthoryear{Grevesse \& Sauval}{1998}]{grev98}
Grevesse, N., \& Sauval, A. J. 1998, {\em Space Science Reviews}, 85, 161

\bibitem[\protect\citeauthoryear{Hauschildt et al.}{1999}]{hau99}
Hauschildt, P. H., Allard, F., Ferguson, J., Baron, E., \& Alexander, D. R. 1999,
ApJ, 525, 871

\bibitem[\protect\citeauthoryear{Hodge}{1967}]{hod67}
  Hodge, P. W., 1967, PASP, 79, 29

\bibitem[\protect\citeauthoryear{Johnson, Bernat \& Krupp}{1980}]{jbk80}
Johnson, H. R., Bernat, A. P., \& Krupp, B. M. 1980, ApJS, 42, 501

\bibitem[\protect\citeauthoryear{Keller}{1999}]{k99}
Keller, S. C. 1999, AJ, 118, 889

\bibitem[\protect\citeauthoryear{Kobulnicky et al.}{1998}]{kob98}
  Kobulnicky, H. A., Kennicutt, R. C., Jr., \& Pizagno, J. L., 1998,
  ApJ, 514, 544

\bibitem[\protect\citeauthoryear{Lan\c{c}on \& Mouhcine}{2000}]{lm00}
  Lan\c{c}on, A., \& Mouhcine, M., 2000, in Massive Stellar Clusters, 
  ASP Conf.\ Ser.\ 211, eds.\ A.\ Lan\c{c}on \* C.\ M.\ Boily, p.\ 34

\bibitem[\protect\citeauthoryear{Larsen \& Richtler}{1999}]{lr99}
  Larsen, S. S., \& Richtler, T., 1999, A\&A 345, 59

\bibitem[\protect\citeauthoryear{Larsen et al.}{2001}]{laretal01}
  Larsen, S. S., Brodie, J. P., Elmegreen, B. G., Efremov, Yu. N., 
  Hodge, P. W. and Richtler, T., 2001, ApJ, 556, 801

\bibitem[\protect\citeauthoryear{Larsen et al.}{2002}]{lar02}
  Larsen, S. S., Efremov, Y. N., Elmegreen, B. G., Alfaro, E. J., 
  Battinelli, P., Hodge, P. W., \& Richtler, T. 2002, ApJ, 567, 896

\bibitem[\protect\citeauthoryear{Larsen et al.}{2004}]{lbh04}
  Larsen, S. S., Brodie, J. P., \& Hunter, D. A., 2004, AJ, 128, 2295

\bibitem[\protect\citeauthoryear{Luck}{1994}]{luck94}
  Luck, R.\ E., 1994, ApJ Suppl., 91, 309

\bibitem[\protect\citeauthoryear{Maeder \& Mermilliod}{1981}]{mm81}
  Maeder, A., \& Mermilliod, J.-C., 1981, A\&A, 93, 136

\bibitem[\protect\citeauthoryear{Massey \& Olsen}{2003}]{mas03}
Massey, P., \& Olsen, K.A.G. 2003, ApJ, 126, 2867

\bibitem[\protect\citeauthoryear{McLean et al.}{1998}]{ml98}
McLean, I. et al. 1998, SPIE, 3354, 566

\bibitem[\protect\citeauthoryear{McWilliam}{1997}]{mac97}
McWilliam, A., 1997, ARA\&A, 35, 503

\bibitem[\protect\citeauthoryear{Mel\'endez \& Barbuy}{1999}]{mb99}
Mel\'endez, J., \& Barbuy, B. 1999, ApJS, 124, 527

\bibitem[\protect\citeauthoryear{Oliva \& Origlia}{1998}]{oo98}
Oliva, E., \& Origlia, L. 1998, A\&A, 332, 46

\bibitem[\protect\citeauthoryear{Origlia, Moorwood \& Oliva}{1993}]{ori93}
Origlia, L., Moorwood, A. F. M., \& Oliva, E. 1993, A\&A, 280, 536

\bibitem[\protect\citeauthoryear{Origlia et al.}{1997}]{ori97}
Origlia, L., Ferraro, F. R., Fusi Pecci, F., \& Oliva, E. 1997, A\&A, 321, 859

\bibitem[\protect\citeauthoryear{Origlia et al.}{1999}]{o99}
Origlia, L., Goldader, J. D., Leitherer, C., Schaerer, D., Oliva, E. 1999, ApJ, 514, 96

\bibitem[\protect\citeauthoryear{Origlia \& Oliva}{2000}]{oo00}
Origlia, L., \& Oliva, E. 2000, NAR, 44, 257

\bibitem[\protect\citeauthoryear{Origlia, Rich \& Castro}{2002}]{ori02}
Origlia, L., Rich, R. M., \& Castro, S. 2002, AJ, 123, 1559 

\bibitem[\protect\citeauthoryear{Origlia et al.}{2003}]{ori03}
Origlia, L., Ferraro, F. R., Bellazzini, M. \& Pancino, E. 2003, ApJ, 591, 916

\bibitem[\protect\citeauthoryear{Origlia et al.}{2004}]{ori04}
Origlia, L., Ranalli, P., Comastri, A., Maiolino, R. 2004, ApJ, 127, 3422

\bibitem[\protect\citeauthoryear{Renzini \& Fusi Pecci}{1988}]{ren88}
Renzini, A., \& Fusi Pecci, F., 1988, ARA\&A, 26, 199

\bibitem[\protect\citeauthoryear{Rich \& Origlia}{2005}]{ro05}
Rich, R. M., \& Origlia, L. 2005, ApJ, in press 

\bibitem[\protect\citeauthoryear{Schaller et al.}{1992}]{schal92}
Schaller, G., Schaerer, D., Meynet, G., \& Maeder, A., 1992, A\&AS, 96, 269

\bibitem[\protect\citeauthoryear{Stasi{\'n}ska}{2001}]{sta01}
Stasi{\'n}ska, G., 2001, Ap\&SS, Suppl., 1, 277, 189

\bibitem[\protect\citeauthoryear{Trager et al.}{1998}]{tra98}
Trager, S. C., Worthey, G., Faber, S. M., Burstein, D., Gonz{\'a}lez, J. J.,
1998, ApJ Suppl., 116, 1

\bibitem[\protect\citeauthoryear{Vazdekis et al.}{2003}]{vaz03}
Vazdekis, A., Cenarro, A. J., Gorgas, J., Cardiel, N., \& Peletier, R. F.,
  2003, MNRAS, 340, 1317

\end{thebibliography}
\end{document}